# Long-distance excitation of nitrogen-vacancy centers in diamond via surface spin waves

Daisuke Kikuchi[1], Dwi Prananto[1], Kunitaka Hayashi[1], Abdelghani Laraoui[2,‡], Norikazu Mizuochi[3,4], Mutsuko Hatano[4,5], Eiji Saitoh[6,7,8,9], Yousoo Kim[10], Carlos A. Meriles[2,11], and Toshu An[1,*]

[1]*School of Materials Science, Japan Advanced Institute of Science and Technology, Nomi, Ishikawa, 923-1292, Japan*
[2]*Dept. of Physics, CUNY-City College of New York, New York, NY, 10031, USA*
[3]*Institute for Chemical Research, Kyoto University, Uji, Kyoto, 610-0011, Japan*
[4]*CREST JST, Kawaguchi, Saitama, 332-0012, Japan*
[5]*Tokyo Institute of Technology, Meguro, Tokyo, 152-8550, Japan*
[6]*Institute for Materials Research, Tohoku University, Sendai, Miyagi, 980-8577, Japan*
[7]*WPI Advanced Institute for Materials Research, Tohoku University, Sendai, Miyagi, 980-8577, Japan*
[8]*Center for Spintronics Research Network, Tohoku University, Sendai, Miyagi, 980-8577, Japan*
[9]*Advanced Science Research Center, Japan Atomic Energy Agency, Tokai, Ibaraki, 319-1195, Japan*
[10]*RIKEN, Hirosawa, Wako, Saitama, 351-0198, Japan*
[11]*CUNY-Graduate Center, New York, NY, 10016, USA*

Coherent communication over mesoscale distances is a necessary condition for the application of solid-state spin qubits to scalable quantum information processing. Among other routes under study, one possibility entails the generation of magnetostatic surface spin waves (MSSW) dipolarly coupled to shallow paramagnetic defects in wide-bandgap semiconductors. As an initial step in this direction, here we make use of room-temperature MSSWs to mediate the interaction between the microwave field from an antenna and the spin of a nitrogen-vacancy (NV) center in diamond. We show that this transport spans distances exceeding 3 mm, a manifestation of the MSSW robustness and long diffusion length. Using the NV spin as a local sensor, we find that the MSSW amplitude grows linearly with the applied microwave power, suggesting this approach could be extended to amplify the signal from neighboring spin qubits by several orders of magnitude.

The realization of chip-integrated, spin-based quantum information processing devices depends on the ability to controllably link distant spin qubits via a coherent quantum bus. Architectures based on linear chains of paramagnetic point defects in wide-bandgap semiconductors have been proposed as an alternative[1,2], but the short range of the dipolar interaction between neighbors and disorder in their relative positions impose engineering challenges presently difficult to overcome. Flying qubits in the form of emitted photons circumvent many of these problems, though entanglement with the spin of the source defect is possible only under cryogenic conditions, and the required spin-photon interfaces are comparatively bulky[3]. Other proposed routes to solid-state qubit communication include the use of photo-ionized carriers electrostatically guided from source to target spin clusters[4], coherent tunneling from neighboring paramagnetic point defects[5], and phonon-mediated coupling[6].

Spin waves in ferrimagnetic materials are presently being explored for classical data transport and information storage as well as for logical operations[7-11]. The advantage of such spintronics and magnonic devices stems for the comparatively low energy dissipation during transport. Whether spin waves in close proximity to paramagnetic defects can serve as a coherent, inter-spin-qubit bus is an intriguing complementary question, which, nonetheless, has received comparatively less attention. The problem has been investigated theoretically and a recent proposal suggests that gapped ferromagnets could coherently mediate the interaction between point defects to produce entanglement over µm-long distances under ambient conditions[12]. Similar mechanisms could also prove useful for nanoscale metrology[13], all the more appealing given the near-matching condition between the resonance frequencies of paramagnetic defects at low magnetic fields and common spin wave modes.

In line with these ideas, here we study the response of nitrogen-vacancy (NV) centers in diamond to magnetostatic surface spin waves (MSSW) propagating in an yttrium iron garnet (YIG) disk. Using a microwave (MW) antenna to excite spin waves in the ferrimagnetic material, we demonstrate MSSW-mediated coherent manipulation of NVs at distances exceeding 3 mm from the MW source. Comparing the MSSW and MW field amplitudes at the NV site, we determine enhancement factors of up to two orders of magnitude. We find no lower bound in the amplitude of the microwave required to inject MSSWs into the disk and a linear response in the limit of low MW fields, thus supporting the idea of ferromagnet-enhanced nanoscale magnetometry and spin sensing[13]. These results complement prior observations in an NV-on-YIG system studied under off-resonance excitation[14,15,16]. During the writing of this manuscript, we learned that experiments similar to ours have been carried out recently with comparable outcome[17].

---

‡ Present Address: *Center for High Technology Materials and Dept. of Physics and Astronomy, University of New Mexico, Albuquerque, NM 87106, USA.*

\* Corresponding author: toshuan@jaist.ac.jp

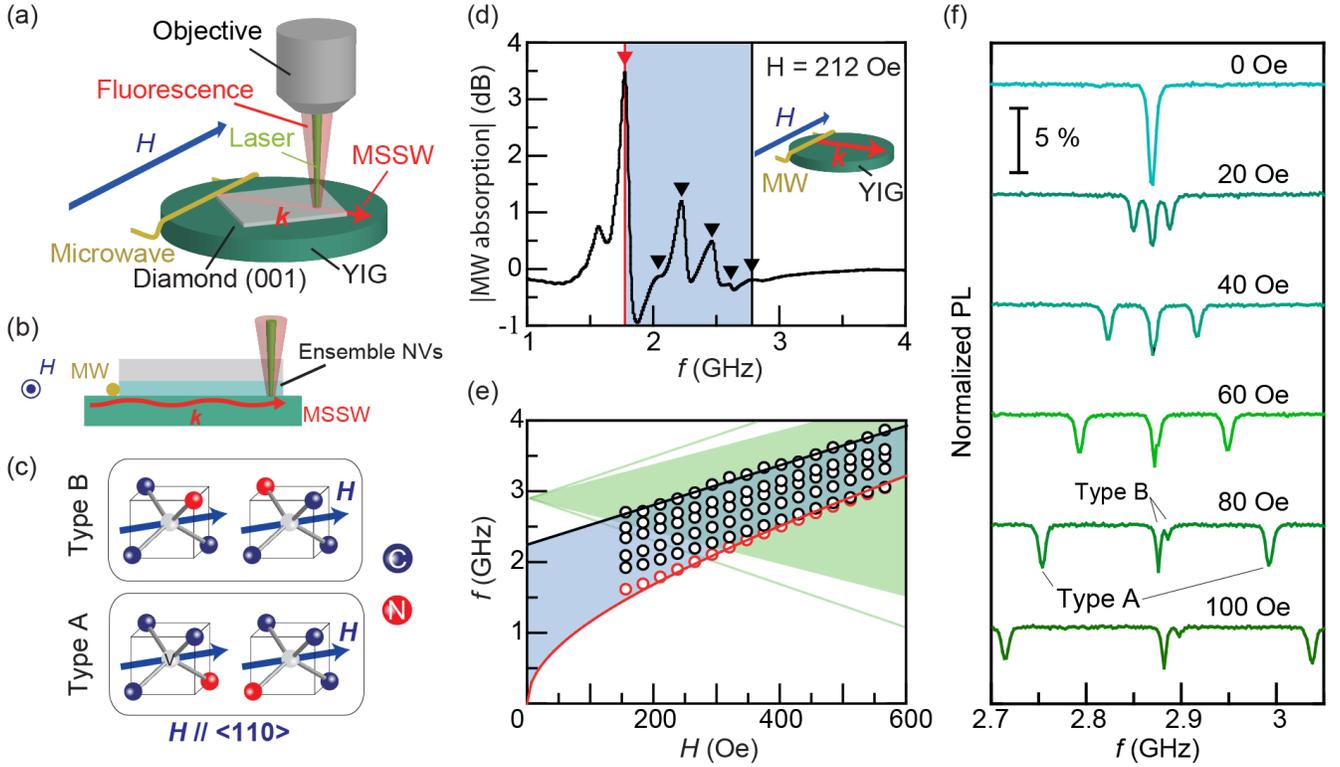

**Figure 1:** (a) Experimental setup. A 532 nm laser illuminates an NV-hosting diamond sitting on a YIG disk, and the resultant fluorescence is collected via a high-NA objective. (b) Cross-sectional view of the cartoon in (a). Microwave and magnetic ($H$) fields are applied to excite MSSWs in the YIG disk along the axis perpendicular to the wire. (c) In the present geometry, we distinguish two NV types, A and B, depending on the projection their axes have along the magnetic field direction. (d) Microwave absorption spectrum for the YIG disk at 212 Oe. The red triangle shows the FMR mode, whereas the blue background indicates the range of MSSW excitation. (e) Dispersion of the microwave absorption signal (contained within the light blue background). Red circles mark the position of the FMR mode at different fields. The black circles indicate the MSSW peaks. The solid green lines are the $|m_S = 0\rangle \rightarrow |m_S = \pm 1\rangle$ transitions for NV⁻s aligned with $H$; the area in light green approximately corresponds to the range of NV⁻ frequencies for our experimental conditions. (f) NV⁻ ODMR spectra at different magnetic fields in the absence of YIG. Outer (inner) dips correspond to Type A (Type B) spin. The vertical scale bar indicates 5% contrast.

Our experimental setup is sketched in Figure 1(a): We generate MSSWs via a MW-driven, thin gold wire overlaid on a ferrimagnetic substrate formed by a 0.4-mm-thick, 4-mm-diameter polycrystalline disk of $Y_3Fe_5O_{12}$, a ferrimagnetic insulator commonly referred to as yttrium iron garnet or YIG. This YIG disk — already utilized in a previous demonstration of spin-wave-mediated directional heat flow[18] — is well suited for the present experiments because it is known to support MSSWs with long diffusion lengths [19,20]. To probe the interaction between the MSSW and insulator-hosted spin qubits, we use NV⁻ centers near the surface of diamond. Formed by a substitutional nitrogen and an adjacent vacancy, the NV⁻ is a spin-1 defect amenable to optical spin initialization and readout[21]. For the present application, we use a <100> type-IIa diamond crystal hosting 30-40 nm deep NVs created via ion implantation (30 keV at a dose of $2.0 \times 10^{11}$ ions/cm²) and annealing (1h at 1000 °C in an argon atmosphere). The NV concentration is estimated at 0.6 ppm, yielding an average inter-NV distance of 60 nm; non-linear responses due to NV-NV interactions can therefore be safely ignored. With the implanted surface of the diamond facing the YIG disk, NVs are illuminated via a 532 nm laser focused to an ~800 nm spot via a high-numerical-aperture objective, itself a part of a customized confocal microscope. Figure 1(b) shows a cross-sectional view of the working geometry: To excite MSSWs propagating on the YIG's surface, we position the Au wire directly on the ferrimagnetic disk and next to one of the diamond crystal corners. An external magnetic field, $H$, is aligned collinear with both the Au wire and the <110> direction of the diamond crystal. In this configuration, MSSWs propagate perpendicular to the external magnetic field[22], and the four possible NV axes group to form two pairs of equivalent directions (type A and B spins in Figure 1(c)).

For our experiments we first determine the YIG response to microwave excitation of variable frequency. Figure 1(d) shows the YIG absorption spectrum for a magnetic field of 212 Oe: In addition to the ferromagnetic resonance (FMR) peak at 1.8 GHz (indicated via a red triangle), we identify five different MSSW modes within the range $f = 2.0$-$2.8$ GHz (black triangles). Figure 1(e) shows the observed central frequencies of each of these resonances as a function of the applied magnetic field $H$. The solid red line is the calculated FMR mode dispersion using Kittel's formula[23], whereas the blue line is a fit using the expected MSSW dispersion relation[18,24]. Also plotted in this graph is the range of resonance frequencies for NV⁻ spins exposed to variable magnetic fields (light green background). Starting from a zero-field splitting of 2.87 GHz, the magnetic field breaks the degeneracy between the $|m_S = 0\rangle \rightarrow |m_S = \pm 1\rangle$ transitions of the NV$^-$ ground triplet ($^3A_2$) thus resulting in two resonances, whose exact values depend on the orientation of the NV axis relative to the applied magnetic field[25]. The two extremes (solid green lines) correspond to NVs collinear with $H$; for our present experimental conditions, however, none of the four possible NV directions coincide with the applied field, which effectively reduces the Zeeman splitting to a narrower range (light green area, see also Figure 1(f)). Importantly, these spin resonances partly overlap with some of the MSSW modes, hence making it possible for the NVs and spin waves to resonantly exchange energy, as we show next.

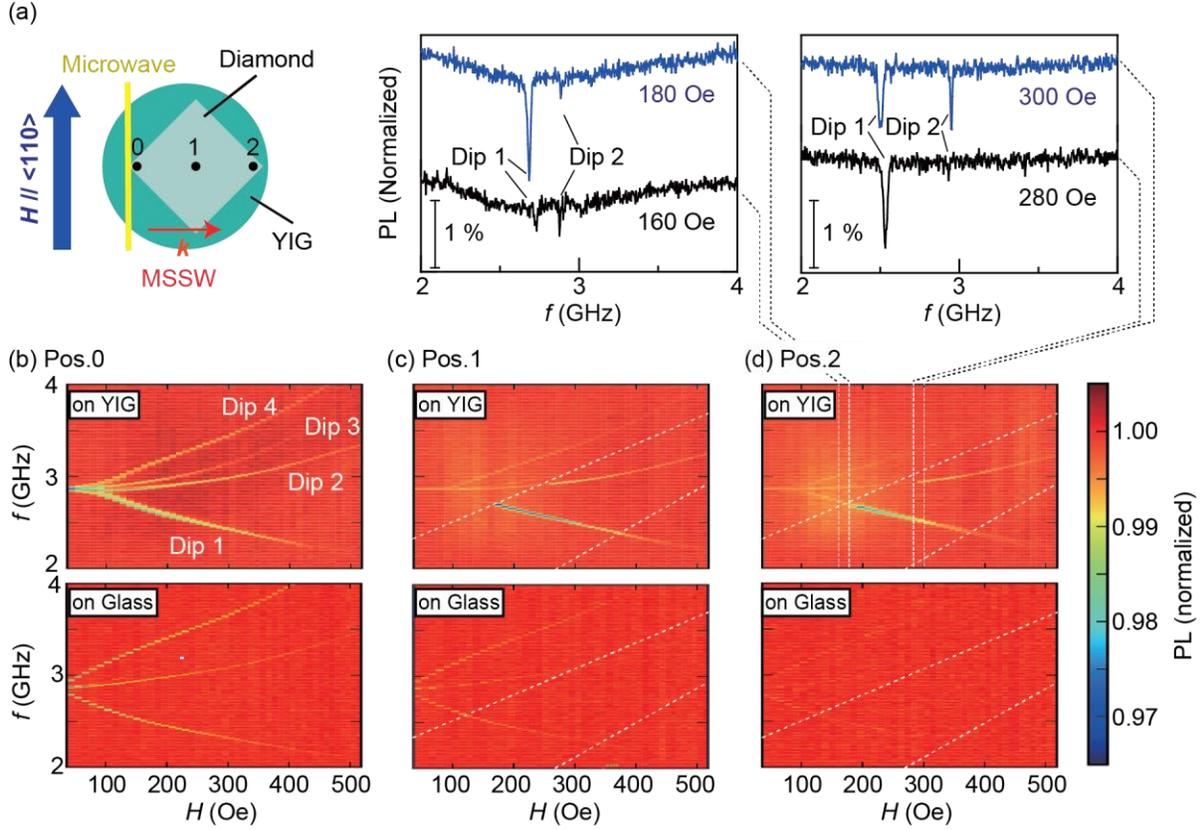

**Figure 2:** (a) Top view of the experimental setup. Solid circles indicate three measurement locations, either immediately next to the gold wire (Position 0), or farther to the right, at 1.8 mm (Position 1), and 3.6 mm (Position 2). (b-d) ODMR spectra as a function of the applied magnetic field for Positions 0, 1, and 2, respectively. Dips 1 and 4 (2 and 3) correspond to spin resonances in Type A (Type B) NVs as presented in Figure 1(c). For comparison, the lower graphs show the results from identical measurements in the case where the diamond crystal sits directly on a glass substrate, i.e., no YIG disk is present. The dashed lines indicate the MSSW frequency range. The upper inserts in (d) show cross sections of the main plot at different magnetic fields (dotted vertical lines). In (b) through (d) the MW power is 560 mW.

A characterization of the NV response under the combined action of spin waves and the MW field is presented in Figure 2. Here we measure the NV$^-$ fluorescence as we sweep the microwave frequency for different applied magnetic fields. The result is a collection of optically detected magnetic resonance (ODMR) spectra exhibiting fluorescence dips at the NV$^-$ spin transition frequencies [26]. We carry out identical sets of measurements at three different locations in the diamond crystal increasingly farther from the MW source (Positions 0, 1, and 2 in Figure 2(a)). Immediately next to the wire (upper graph in Figure 2(b)) we distinguish four spin resonances (Dips 1 through 4), which we associate with the $|m_S = 0\rangle \rightarrow |m_S = \pm 1\rangle$ transitions of the two non-equivalent NV directions present in the chosen working geometry (Figure 1(c)). MW-field-induced ODMR dips are also observable when we replace the ferrimagnetic disk by a glass cover slip (lower graph in Figure 2(b)), though the signal amplitudes are smaller to the point of making one of the four dips disappear. The effect is even more dramatic at greater distances from the wire where virtually no ODMR signal is present if the YIG disk is not present (Figs. 2(c) and 2(d)).

We interpret the above results in terms of an energy exchange between the NVs and the MSSWs propagating in the YIG disk. The observations in Fig. 2 strongly support this notion: For example, considering the group of NVs with the largest projection along $H$ (Type A in the insert to Figure 1(c)), we find that the largest ODMR dips are observed near energy matching conditions, when one of the NV spin resonances coincides with the frequency of a spin wave mode. Further, in Figure 2(d) the nearly featureless spectrum observed at 160 Oe abruptly changes at 180 Oe to exhibit a strong fluorescence dip at ~2.7 GHz, when the $|m_S = 0\rangle \rightarrow |m_S = -1\rangle$ transition matches the MSSW mode of highest frequency. Note that Dip 4 — corresponding to the $|m_S = 0\rangle \rightarrow |m_S = +1\rangle$ transition for type A NVs — is much weaker because the YIG disk supports nearly no MSSWs in this frequency range (see Figure 1(e)). We witness a comparable change at higher magnetic fields, when the spin wave modes overlap with the $|m_S = 0\rangle \rightarrow |m_S = -1\rangle$ transition in NVs of type B (oriented perpendicular to the external magnetic field, Figure 1(c)); as before, only one of the resonances is enhanced (compare spectra at 280 Oe and 300 Oe in the right insert to Figure 2(d)).

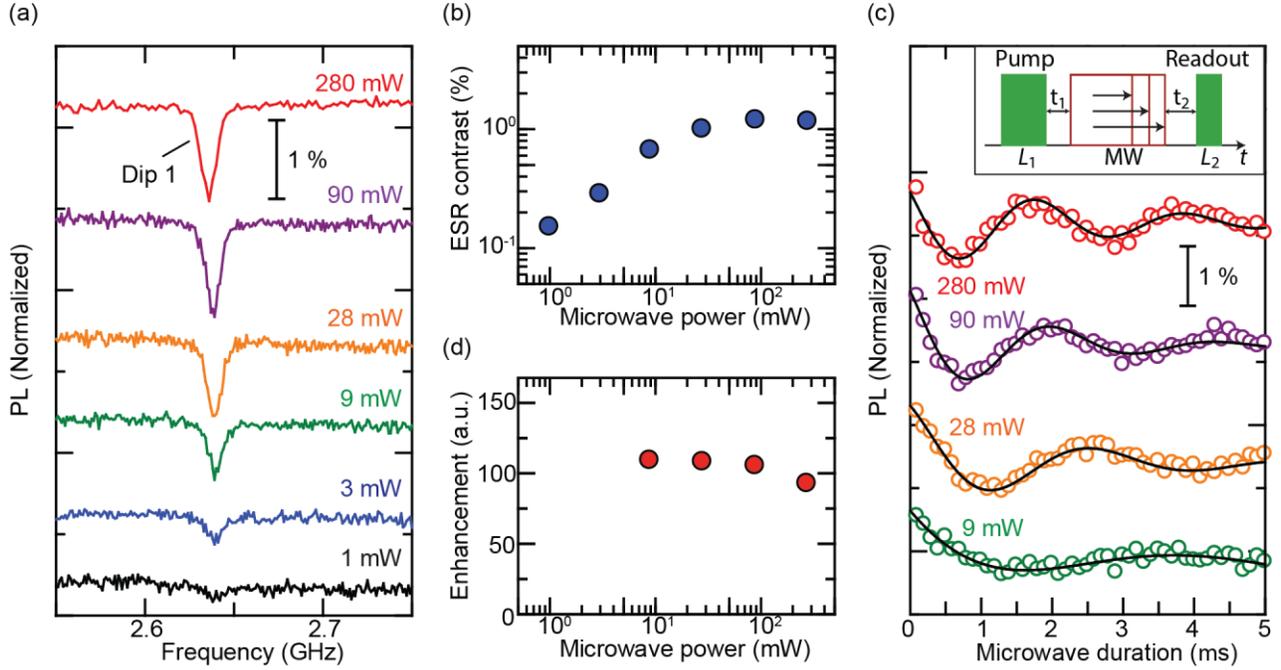

**Figure 3:** (a) NV⁻ ODMR spectra in a frequency range around Dip 1 at different MW powers. Spectra have been displaced vertically for clarity. (b) ODMR dip amplitude as a function of the MW power. (c) Rabi oscillations of the $|m_S = 0\rangle \rightarrow |m_S = -1\rangle$ transition in type A NV spins (Dip 1). The laser power is 0.53 mW and the laser pulse durations are 3 µs and 0.5 µs for $L_1$ and $L_2$, respectively. To mitigate spurious transients in our electronics we introduce two time delays, $\tau_1$=30 µs and $\tau_2$=2 µs, following the pump and MW pulses, respectively. The vertical scale bar indicates 1% contrast. (d) Enhancement of the Rabi field determined as the ratio between the Rabi frequency observed in (c) and the Rabi frequency when no YIG is present. The latter is calculated from extrapolating the value observed at 80 µm from the wire and assuming a dependence inversely proportional to distance. In (a) through (d) the magnetic field is $H$=212 Oe and the distance to the antenna is 3.6 mm.

Figure 3(a) shows a series of ODMR spectra at different MW powers; in these experiments the distance to the MW source is approximately 3.6 mm and the external magnetic field is $H$=212 Oe. Plotting the ODMR dip amplitude as a function of the microwave power, we find a linear response though saturation is present beyond ~10 mW. The latter is a consequence of the reduced ODMR contrast, limited to about 1% at these magnetic fields (see Figure 1(f)). To demonstrate time control of the NV⁻ spin, we gradually increase the duration of a MW pulse while keeping the power constant; the resulting Rabi oscillations for four different microwave powers are shown in Figure 3(c). For 90 mW, we measure a Rabi period $T_R$~2 µs and correspondingly deduce an effective Rabi field of ~0.17 Gauss, approximately ~100 times larger than the calculated MW field at this same position when no YIG disk is present (Figure 3(d)). At higher powers we observe slight saturation of the NV⁻ spin Rabi frequency and thus a reduction of the enhancement, possibly due to a non-linear response of the YIG disk. Additional work, however, will be needed to fully understand this behavior.

In summary, our work shows that surface spin waves in a ferrimagnetic material can effectively mediate the interaction between a solid-state spin qubit and a MW antenna under ambient conditions. Using a YIG disk in direct contact with a diamond crystal we demonstrate time-resolved, spin-wave-enabled manipulation of NV⁻ centers far removed from the antenna, in regions where the MW field is negligible. This process is resonant and takes place when the frequency of an MSSW mode matches the transition frequency between spin levels in the paramagnetic center. Comparing the spin wave and antenna dipolar fields near the probe NVs we determine an enhancement exceeding one hundred fold. This amplification is, of course, a function of the distance to the MW source, the MSSW diffusion length, and the system geometry.

An intriguing question is whether the above mechanism can be effective in mediating the dipolar interaction between two spin-qubits too far from each other to couple directly. This problem is relevant not only to quantum information processing but, more broadly, to nanoscale spin sensing where a probe spin (such as the NV⁻ center) must detect the signal from another spin moiety outside the diamond crystal. The development of schemes to actively control the spin-wave field will be key to the implementation of these ideas. On the other and, the ability to operate with deeper sensor spins (tens of nanometers from the host surface) could prove advantageous given the deleterious effects of dangling bonds and other surface paramagnetic defects.

## ACKNOWLEDGEMENTS

D.K. and T.A. acknowledge support from JSPS KAKENHI Grant (No. JP16H06831), Grant-in-Aid for Scientific Research on Innovative Area, "Nano Spin Conversion Science" (Grant No. 26103002), The Murata Science Foundation, and The Canon Foundation. N.M. is supported from CREST, JST, and JSPS KAKENHI Grants (No. 15H05868, 16H02088). M.H. is supported by JSPS KAKENHI Grant (No. 26820110) and JST-CREST. E.S. acknowledges support from Grant-in-Aid for Scientific Research on Innovative Area, "Nano Spin Conversion Science" (Grant No. 26103006), ERATO "Spin Quantum Rectification Project", Japan Grant-in-Aid for scientific research (A) (No. 15H02012) from MEXT, Japan. Y. K. acknowledges support by Grant-in-Aid for Scientific Research (A) (No.15H02025). A.L. and C.A.M. acknowledge support from the US National Science Foundation under grants NSF-1619896 and NSF-1401632 as well as Research Corporation through a FRED Award.